\documentclass[sigconf,9pt]{acmart}
\usepackage{amsmath}
\usepackage{mathrsfs}
\usepackage{multirow}
\usepackage{multicol}
\usepackage{pifont}
\usepackage{bbding}
\usepackage{placeins}
\usepackage{rotating}
\usepackage{setspace}
\usepackage{soul}
\usepackage[hang, tight]{subfigure}
\usepackage{threeparttable}
\usepackage{url}
\usepackage{upgreek}
\usepackage{verbatim}
\usepackage{wrapfig}
\usepackage[]{siunitx}
\usepackage[super]{nth}
\usepackage{tabularx}
\usepackage{supertabular}
\usepackage{tikz}
\usepackage{bbm}
\usepackage{blindtext}
\usepackage{makecell}
\usepackage{adjustbox}
\usepackage{afterpage}
\usepackage{csquotes}
\usepackage{booktabs}
\usepackage{balance}

\usepackage{algpseudocode}
\algtext*{EndWhile}
\algtext*{EndIf}
\algtext*{EndFor}
\algtext*{EndProcedure}

\graphicspath{{./_fig/}}

\makeatletter
\renewcommand\footnoterule{%
  \kern-3\p@
  \hrule\@width0.4\columnwidth
  \kern2.6\p@}
  \makeatother

\usepackage{graphicx}
\usepackage{adjustbox}

\def\fullcheckmark{\tikz\draw[scale=0.4,fill=black](0,.35) -- (.25,0) -- (1,.7) -- (.25,.15);}

\newcommand*\circled[1]{\tikz[baseline=(char.base)]{
            \node[shape=circle,fill,inner sep=1pt,scale=0.8] (char) {\textcolor{white}{#1}};}}
            
\usepackage{bbding}

\usepackage{float}

\usepackage{color}

\definecolor{gg}{RGB}{0, 129, 35}
\definecolor{orangeShallow}{RGB}{255,190,0}

\usepackage[english]{babel}
\usepackage[utf8]{inputenc}
\usepackage{tabulary}
\usepackage{colortbl}
\usepackage{cleveref}
\copyrightyear{2025} 
\acmYear{2025} 
\setcopyright{acmlicensed}\acmConference[ASPDAC '25]{30th Asia and South
Pacific Design Automation Conference}{January 20--23, 2025}{Tokyo, Japan}
\acmBooktitle{30th Asia and South Pacific Design Automation Conference
(ASPDAC '25), January 20--23, 2025, Tokyo, Japan}
\acmDOI{10.1145/3658617.3697756}
\acmISBN{979-8-4007-0635-6/25/01}

\begin{document}







\title{AssertLLM: Generating Hardware Verification Assertions from Design Specifications via Multi-LLMs}











\author{Zhiyuan Yan$^{1*}$, Wenji Fang$^{2*}$, Mengming Li$^2$, Min Li$^3$, Shang Liu$^2$, Zhiyao Xie$^{2\textsuperscript{†}}$, Hongce Zhang$^{1,2\textsuperscript{†}}$}

\affiliation{%
  \institution{$^1$Hong Kong University of Science and Technology (Guangzhou)}
  \country{}
}
\affiliation{%
  \institution{$^2$Hong Kong University of Science and Technology}
  \country{}
}
\affiliation{%
  \institution{$^3$Huawei Technologies Co., Ltd.}
  \country{}
}

\thanks{$^*$These authors contributed equally to this work.}
\thanks{$\textsuperscript{†}$Corresponding Authors: \{hongcezh, eezhiyao\}@ust.hk}

\begin{abstract}
Assertion-based verification (ABV) is a critical method to ensure logic designs comply with their architectural specifications. ABV requires assertions, which are generally converted from specifications through human interpretation by verification engineers. Existing methods for generating assertions from specification documents are limited to sentences extracted by engineers, discouraging their practical applications. 
In this work, we present AssertLLM, an automatic assertion generation framework that processes complete specification documents. AssertLLM can generate assertions from both natural language and waveform diagrams in specification files. It first converts unstructured specification sentences and waveforms into structured descriptions using natural language templates. Then, a customized Large Language Model (LLM) generates the final assertions based on these descriptions. 
Our evaluation demonstrates that AssertLLM can generate more accurate and higher-quality assertions compared to GPT-4o and GPT-3.5. \looseness=-1

\end{abstract}

\maketitle

\pagestyle{empty}

\section{Introduction}\label{sec:intro}

Hardware functional verification is critical in the VLSI design flow. It addresses the following question: whether an implementation adheres to its specification. 
The specifications are typically drafted in natural language by architects and then translated into RTL code by designers. Verification engineers then check the functional correctness of the RTL designs according to the specifications.
In the verification process, assertion-based verification (ABV)~\cite{witharana2022survey} is a widely adopted technique, which utilizes assertions crafted from specifications to verify the functional behavior of RTL designs. ABV can be performed either by simulation or formal property verification (FPV), where assertions are often written in the form of SystemVerilog Assertions (SVAs). However, a significant challenge in ABV is the generation of sufficient, high-quality assertions. Currently, designing SVAs manually is a time-consuming and error-prone task, demanding unignorable human effort.

To address this challenge, research has focused on generating SVAs automatically, which can be mainly categorized into two types: (1) dynamic mining from simulation traces and (2) static analysis of specifications. 
Dynamic methods~\cite{germiniani2022harm, danese2017team, vasudevan2010goldmine} combine simulation with static design constraint analysis but risk generating incorrect SVAs due to their reliance on potentially flawed RTL designs.
Static methods utilize predefined templates~\cite{orenes2021autosva, fang2023r} or natural language processing (NLP)-based machine learning (ML) technologies~\cite{harris2016glast,krishnamurthy2019controlled,zhao2019automatic,krishnamurthy2019ease,frederiksen2020automated,keszocze2019chatbot,parthasarathy2021spectosva,aditi2022hybrid, chang2024data}. Recently, the potential of generative AI technologies like Large Language Models (LLMs) has gained significant attention in hardware design process~\cite{lu2024rtllm, liu2024rtlcoder, chen2024dawn, liu2023verilogeval, li2024specllm, pei2024betterv, wu2024chateda}. Researchers are also exploring the use of LLMs to generate hardware assertions~\cite{kande2023llm, orenes2023using, sun2023towards, pulavarthi2024assertionbenchbench, liu2024openllm}.


We further categorize the existing static ML-based methods based on their application in different design phases: the RTL and pre-RTL stages.
Table~\ref{tbl:related_work} details these ML-based SVA generation methods in both the RTL stage and the pre-RTL stage. During the RTL stage, the process typically involves using LLMs to process both human-written specification sentences and the RTL design to generate SVAs describing security or functional properties~\cite{kande2023llm, liu2024domain, orenes2023using, sun2023towards}. However, similar to the dynamic methods, inaccuracies in RTL implementations could result in flawed SVAs. 



\begin{table}[!t]
\caption{
Existing works on generating SVAs from natural language specifications. AssertLLM is the first work that can handle full-size specification files and generate comprehensive types of SVAs for each architectural signal.
}
\resizebox{0.48\textwidth}{!}{
\begin{tabular}{c||c|c|c|c|c|c}
\toprule
\multirow{2}{*}{\textbf{Stage}} & \multirow{2}{*}{\textbf{Works}}                                                                                                                                                    & \multirow{2}{*}{\textbf{\begin{tabular}[c]{@{}c@{}}Gen. \\      Method\end{tabular}}} & \multicolumn{3}{c|}{\textbf{NL Specification}}                                                                                                         & \multirow{2}{*}{\textbf{\begin{tabular}[c]{@{}c@{}}Verification \\      Target\end{tabular}}}         \\ \cline{4-6}
                                &                                                                                                                                                                                    &                                                                                       & Auto Proc.                    & Source                                                                                & Full Design                   &                                                                                                       \\ \hline \hline
\multirow{2}{*}{RTL}            & \cite{kande2024security, mali2024chiraag}                                                                                                                                                                & \multirow{2}{*}{LLM-based}                                                            & \multirow{2}{*}{\XSolidBrush} & \multirow{2}{*}{\begin{tabular}[c]{@{}c@{}}\textcolor{red}{Sentences}\\      (Engineers)\end{tabular}} & \multirow{2}{*}{\XSolidBrush} & Security                                                                                              \\ \cline{2-2} \cline{7-7}
                                & \cite{orenes2023using,   sun2023towards,liu2024domain}                                                                                                                             &                                                                                       &                               &                                                                                       &                               & \begin{tabular}[c]{@{}c@{}}Function \\      \textcolor{red}{(few examples)}\end{tabular}                               \\ \hline
\multirow{2}{*}{Pre-RTL}        & \cite{harris2016glast,krishnamurthy2019controlled,zhao2019automatic,krishnamurthy2019ease,frederiksen2020automated,keszocze2019chatbot,parthasarathy2021spectosva,aditi2022hybrid} & NLP-based                                                                             & \XSolidBrush                  & \begin{tabular}[c]{@{}c@{}}\textcolor{red}{Sentences}\\      (SPEC)\end{tabular}                       & \fullcheckmark                & \begin{tabular}[c]{@{}c@{}}Function \\      \textcolor{red}{(property checkers/}\\      \textcolor{red}{artificial cases)}\end{tabular} \\ \cline{2-7} 
                                & \cellcolor[HTML]{C5D9F1}Ours                                                                                                                                                                               & \cellcolor[HTML]{C5D9F1}LLM-based                                                                             & \cellcolor[HTML]{C5D9F1}\fullcheckmark                & \cellcolor[HTML]{C5D9F1}\begin{tabular}[c]{@{}c@{}}\textcolor{gg}{Entire Doc}\\      (SPEC)\end{tabular}                      & \cellcolor[HTML]{C5D9F1}\fullcheckmark                & \cellcolor[HTML]{C5D9F1}\begin{tabular}[c]{@{}c@{}}Function \\      \textcolor{gg}{(general designs)}\end{tabular}          \\ 
                                \bottomrule
\end{tabular}
}

\label{tbl:related_work}
\vspace{-.1in}
\end{table}

\begin{figure*}[!t]
  \centering
  \includegraphics[width=1.0\linewidth]{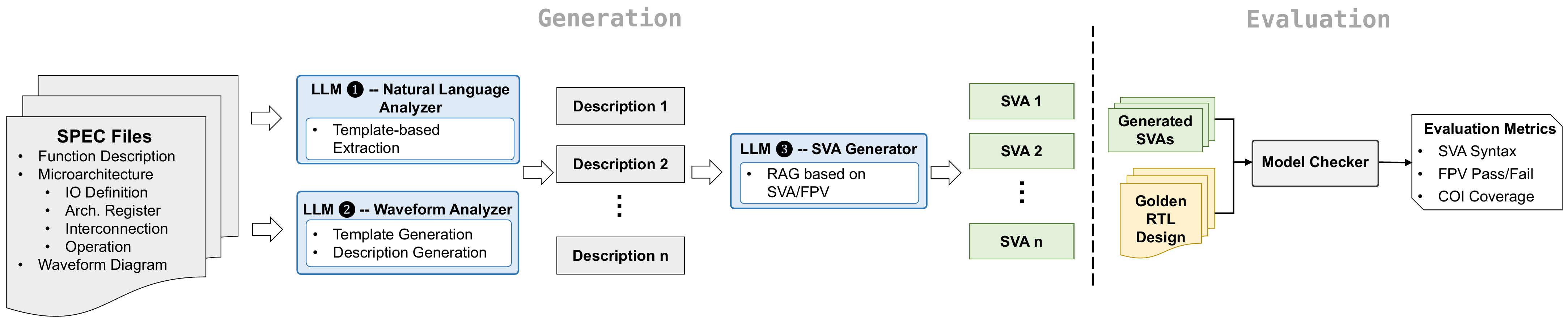}
      \caption{AssertLLM generation and evaluation workflow.
      AssertLLM incorporates three LLMs, each enhanced with specific techniques for the decomposed tasks: extracting structural information from natural language, extracting structural information from waveform diagrams, and translating extractions into various SVA types. The generated SVAs are further evaluated based on the golden RTL implementations using model checking tools.
      }
  \label{fig:workflow}
\end{figure*}

Regarding the pre-RTL stage, with the specification document finalized, RTL designers proceed to implement behavioral designs satisfying the specification.
Numerous studies~\cite{harris2016glast,krishnamurthy2019controlled,zhao2019automatic,krishnamurthy2019ease,frederiksen2020automated,keszocze2019chatbot,parthasarathy2021spectosva,aditi2022hybrid} have used NLP techniques to generate SVAs from natural language sentences, focusing on sentence-level analysis from extensive specifications. However, this approach has several limitations. It is labor-intensive, requiring manual extraction of relevant sentences. It also struggles with diverse grammatical structures, and its evaluation often relies on design-specific checkers, limiting broader applicability. Furthermore, specifications frequently include waveform diagrams that illustrate functional behaviors across different signals. Nevertheless, there is no existing technique that can generate SVAs from these diagrams currently, representing a significant gap in the field.
\looseness=-1

Here we summarize three key challenges that currently hinder the practical application of SVA generation from specification documents:\looseness=-1
\begin{enumerate}
    \item Natural language VLSI specifications are often unstructured and are hard to be directly used for assertion generation.
    \item Even with structured specifications, translating natural language into assertions remains a highly complex task,  requiring both a deep understanding of the design functionality and an expertise in SVA.
    \item While waveforms commonly exist in specification documents, no current research focuses on capturing functional behaviors from waveforms and generating corresponding SVAs. \looseness=-1
\end{enumerate}

To tackle these challenges in SVA generation, in this work, we propose AssertLLM, a novel automatic assertion generation framework incorporating multiple LLMs to deal with the decomposed tasks separately.
This framework is designed to process the entire specification files as they are, and automatically produce SVAs across different signals, significantly benefits both design-time bug prevention and verification-time bug detection. 
AssertLLM operates in three main steps. First, it processes natural language in specification documents, using an LLM to convert unstructured information into a structured format via a unified template. Second, it analyzes waveform diagrams in the specification files, employing another LLM to create natural language templates and extract waveform descriptions based on these templates. Unlike previous methods~\cite{orenes2021autosva, fang2023r}, 
AssertLLM does not require an additional human input 
to create templates. 
Finally, a customized LLM translates the extracted information into SVAs.  The resulting SVAs check various design aspects, including bit-width, connectivity, and functionality.


Our contributions in this work are summarized below:
\begin{itemize}
    \item To the best of our knowledge, AssertLLM is the first automatic assertion generation method that can handle the complete specification files.
    \item We decompose the assertion-generation process into three key steps: extracting structural information from natural language, generating descriptions from waveform diagrams, and translating the information into SVAs. These SVAs support checks for bit-width, connectivity, and functionality.\looseness=-1
    \item To demonstrate the effectiveness of AssertLLM, we conduct a comprehensive evaluation on a set of designs. The result shows that 88\% of generated SVAs are evaluated to be correct both syntactically and functionally. Furthermore, these correct SVAs achieve 97\% cone of influence coverage, demonstrating the high quality of the generated SVAs. 
\end{itemize}

\section{Methodology}

\subsection{Workflow Overview}

Figure~\ref{fig:workflow} illustrates the process of generating and evaluating SVAs for AssertLLM. Our approach integrates three LLMs to generate hardware verification assertions from comprehensive specification documents. These models perform the following tasks: 1) extract relevant information from the natural language in the specification necessary for SVA generation; 2) extract behavioral descriptions from waveform diagrams for SVA generation; 3) generate high-quality SVAs based on the previous extracted information.

In the subsequent subsections, we will detail the functionalities of each LLM of the assertion generation flow. Following this, our SVA evaluation methodology will be presented.\looseness=-1

\subsection{Natural Language Analyzer}\label{sec:nlspec}


A comprehensive natural language specification typically includes seven key sections: 1) introduction: outlines the design's concepts and features; 2) IO ports: provides detailed information for the interface; 3) registers: describes all the architecture-level registers in the design; 4) operation: explains the operational procedures for dataflow and control; 5) architecture: the high-level workflow and dataflow of the design; and 6) usage examples: offers basic usage scenarios for the design.  For signals, the specification may only define critical architecture-level IO ports and registers, leaving the designers to detail internal signals for RTL implementations. 7) waveform diagram: describe behaviors for different signals.

\begin{figure}[!h]
  \centering
  \includegraphics[width=0.82\linewidth]{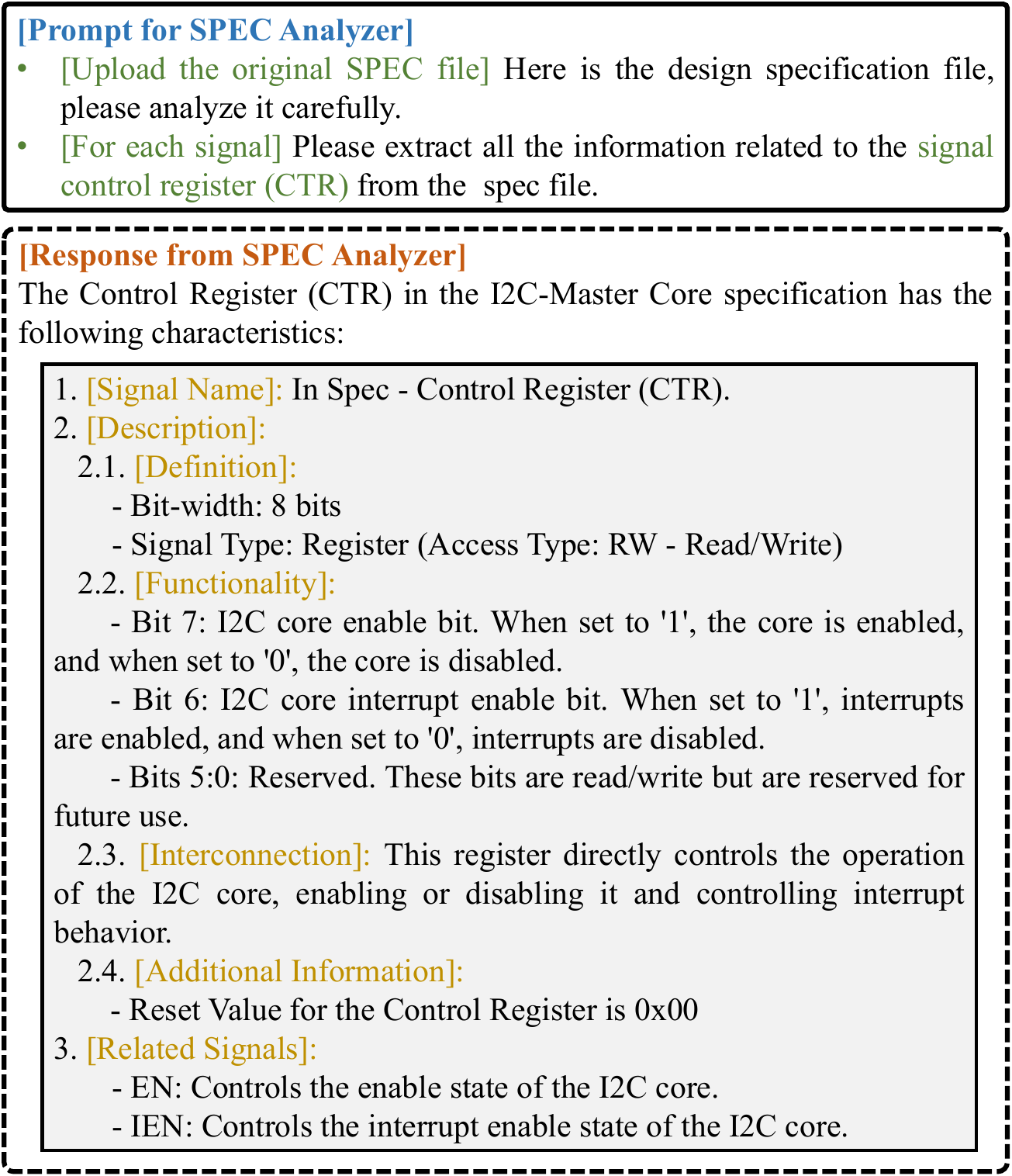}
  \caption{Prompt and Response Example of LLM \protect\circled{1}\protect \texttt{ Natural Language Analyzer}}
  \label{fig:llm1_prompt}
\end{figure}

The first step of our AssertLLM framework is to extract structured information from natural language specification documents to enable SVA generation. 
As we discussed in Section~\ref{sec:intro}, the first key challenge of SVA generation lies in the inherent unstructured nature of the original specifications, which contain background information, functional descriptions, microarchitecture designs, and various diagrams, including dataflow and waveform, etc. Meanwhile, the existence of assertion-relevant information across different sections further complicates the direct utilization of the original specifications for SVA generation.


To address the challenge of processing original, unstructured, full-size specification documents, we utilize an LLM tailored to extract structural and relevant information for each defined signal, thereby further facilitating the SVA generation process.

Specifically, in our LLM \circled{1} \texttt{Natural Language Analyzer}, we first utilize system instructions to customize the LLM.
The model takes the full-size specification file as the input, and the multi-modal function is employed to analyze the file containing text, tables, figures, etc. 
Then for each signal, the LLM is required to extract all the related information of the signal. Here, we design a unified and structured template to guide the LLM in extracting all essential signal-related information. This template contains three key components: the signal's name, its description, and the interconnection signals. We demonstrate the details of each part as follows:
\begin{itemize}
    \item \textbf{Name}: The identifier of the signal in the specification, ensuring clear and unambiguous reference.
    \item \textbf{Description}: To facilitate SVA generation, we divide the descriptions into four categories, including 1) definitions such as bit-width and signal type. 2) functionality which contains all the function-related information of the target signal in the entire specification file. 3) interconnection relationship with all other signals. 4) additional information that is not included in the above three types. 
    \item \textbf{Interconnection Signals}: A list of signals that interact or are associated with the target signal, which are essential for the subsequent assertion generation process.
\end{itemize}

Note that the extracted information is summarized across different sections of the original specification, which contains all the \textit{textual} information needed for assertion generation.
In order to incorporate specifications from the waveform, 
we employ another LLM to extract information from the waveform diagrams as presented in next subsection, thereby enriching the set of SVAs and potentially enhancing their overall quality.\looseness=-1

\subsection{Waveform Analyzer}
In this work, we are interested in extracting design behaviors from waveforms in the specification document, where these waveforms are presented in the form of images (a two-dimensional array of pixels). 
Unlike those prior methods~\cite{vasudevan2010goldmine,germiniani2022harm,danese2017team} that assume a numeric and structured waveform, such as those stored in the Value Change Dump (VCD) format, AssertLLM needs to first interpret the images of waveforms to obtain a description of behaviors. 
Techniques like optical character recognition (OCR)~\cite{mori1999optical} may be used to convert diagrams of a fixed format into text, however, it would be hard to accommodate various waveform styles. Whereas, LLM is advantageous in its flexibility.
Despite that multi-modal LLMs can take images as input and there are prior works like image captioning~\cite{anderson2018bottom,aneja2018convolutional} that interpret input images with human languages. However, they are not suitable for waveform interpretation.

\begin{sloppypar}
To address this problem, we propose another LLM \circled{2} \texttt{Waveform Analyzer} to extract behavior descriptions from waveform diagrams. This analyzer employs natural language to describe the behavior of various signals in the waveform through a two-step process. In the first step,  \texttt{Waveform Analyzer} automatically creates a set of templates suitable for describing the behaviors in the given waveform. Subsequently, \texttt{Waveform Analyzer} produces the behavioral descriptions based on these templates.
\end{sloppypar}
\subsubsection{Template Generation} 
\begin{figure}[t]
  \centering
  \includegraphics[width=0.83\linewidth]{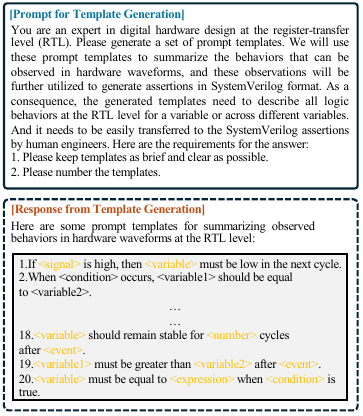}
  \caption{Prompt and Response Example for template generation in LLM \protect\circled{2}\protect~\texttt{Waveform Analyzer}}
  \label{fig:waveform_first_llm_prompt}
  \vspace{-.2in}
\end{figure}

It requires significant effort for a human to write templates to describe behaviors on a single signal or among various signals on the waveform diagrams. To solve this problem, we employ prompt engineering to automate the template generation process. Fig.~\ref{fig:waveform_first_llm_prompt} shows the prompt and corresponding response in the step of template generation. These automatically generated templates serve as references for producing descriptions of given waveform diagrams in the subsequent step.

\begin{sloppypar}
\subsubsection{Description Generation} 
\begin{figure}[!h]
  \centering
  \includegraphics[width=0.99\linewidth]{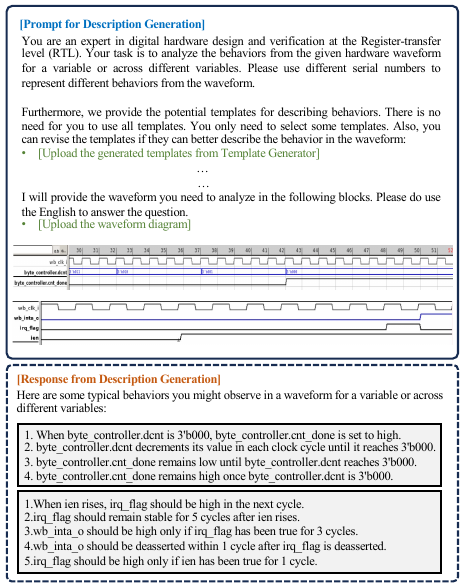}
  \caption{Prompt and Response Example for description generation in LLM \protect\circled{2}\protect~\texttt{Waveform Analyzer}}
  \label{fig:waveform_second_llm_prompt}
\end{figure}

In this step, \texttt{Waveform Analyzer} takes our unified prompt, extracted waveform diagrams, and generated templates as inputs to produce descriptions of the behavior for the given waveform. Fig.~\ref{fig:waveform_second_llm_prompt} shows the corresponding prompt and response in the description generation step. By processing and analyzing the waveform diagrams and the templates, our  \texttt{Waveform Analyzer} comprehensively investigates the functional information in the waveforms, thereby identifying potential relationships among different signals. For example, it can determine whether the output signal is valid when a specific signal is enabled or deduce that a signal should remain stable for a specific number of cycles when the corresponding condition is met. Furthermore, we prompt the LLM to directly use the provided templates or extend them if necessary. This flexibility allows the \texttt{Waveform Analyzer} to explore additional behaviors in the waveform.
\end{sloppypar}

\subsection{Automatic Assertion Generation}

For the conversion from textual descriptions to assertions,
prior research has explored traditional NLP techniques for pre-RTL stages and LLM-based approaches for RTL designs, each with its limitations. NLP methods require detailed syntax and semantics analysis, limiting adaptability to sentence structure variations. LLM approaches, focusing on the RTL stage, depend on HDL code and annotations but risk generating inaccurate SVAs from unverified RTL code, potentially compromising the verification process.\looseness=-1

To address these challenges, our work introduces the LLM \circled{3} \texttt{SVA Generator}, dedicated to generating assertions for each signal utilizing the previously extracted structural information.

\begin{figure}[!h]
  \centering
  \includegraphics[width=0.91\linewidth]{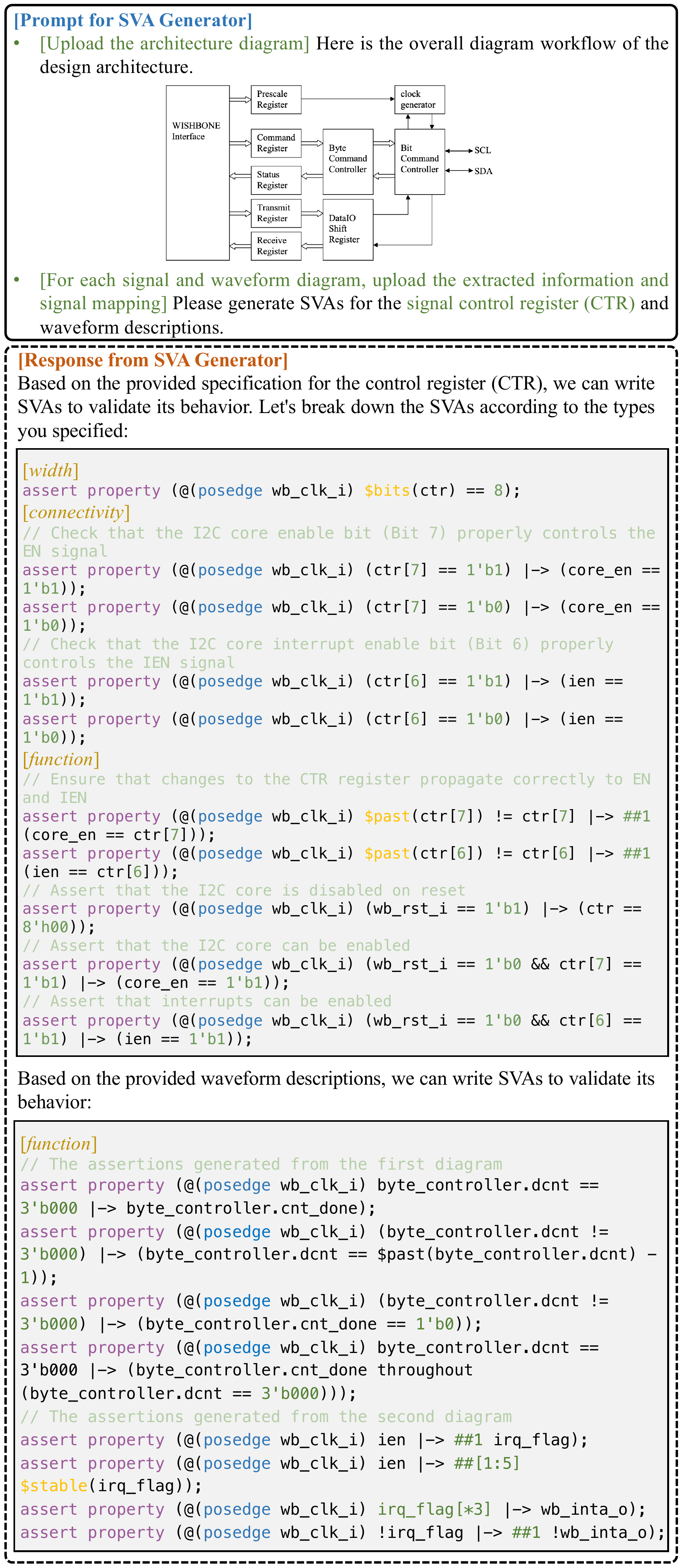}
  \caption{Prompt and Response Example of LLM \protect\circled{3}\protect~\texttt{SVA Generator}}
  \label{fig:llm3_prompt}
  \vspace{-.2in}
\end{figure}

Considering the precise syntax and writing rules inherent to SVAs and the potential for the original LLM failing to generate syntactically correct SVAs, as discussed in~\cite{orenes2023using}, we incorporate the Retrieval Augmented Generation (RAG) technique~\cite{yih2020retrieval} to enhance the LLM's capability for SVA generation. This approach is enriched by a knowledge database comprising tutorials and textbooks on SVA and formal property verification~\cite{seligman2023formal, mehta2020systemverilog, vijayaraghavan2005practical}, providing a robust foundation for the LLM to access and retrieve relevant SVA knowledge based on the input query, thereby enhancing the quality of the generated SVAs.\looseness=-1

\begin{sloppypar}
In addition to the RAG technique, we guide the \texttt{SVA Generator}
with the overall architecture diagram of the design. The LLM is provided with the structured specifications extracted from the previous LLMs for each signal. Then the LLM is required to generate SVAs that adhere strictly to the specifications. 
We categorize SVAs into three types: 1) \textbf{width}: check if the signal bit width satisfies the specification; 2) \textbf{connectivity}: check if the signal can be correctly exercised and also the value propagation among all connected signals; 3) \textbf{function}: check if the function defined in the specification is implemented as expected. Fig.~\ref{fig:llm3_prompt} demonstrates an example of generating SVAs for a signal.   
\end{sloppypar}

\subsection{Evaluation of Generated Assertions}

After the SVA generation process,  we then evaluate the quality of the generated assertions.
While some previous studies such as~\cite{zhao2019automatic, frederiksen2020automated} suggest using specifically-designed checkers for this purpose, such an approach is limited to particular design types like protocols and processors and lacks generalizability to all VLSI designs. Other methods like~\cite{orenes2023using} involve manual inspection by the engineers of FPV results using the generated assertions.
While in our approach, we assume there are golden RTL implementations for the designs. Especially, we pick such designs as test cases as they have been thoroughly tested and can be regarded as  bug-free golden references. \looseness=-1


For evaluation, we also utilize the FPV method. The generated SVAs and the golden RTL designs are fed into a model checker. After performing FPV, we compute the following metrics to evaluate the quality of SVAs: 1) \textbf{syntax}: checks if the generated SVAs have syntax errors; 2) \textbf{FPV pass/fail}: assuming the RTL designs are bug-free, an SVA that passes the FPV check is considered semantically correct, and conversely, a failure indicates an incorrect SVA. 3) \textbf{COI coverage}: Cone of Influence (COI) coverage measures the percentage of design logic that is structurally connected to the properties. It is a common metric to evaluate the quality and usefulness of the generated properties.

\section{Case Study}

\subsection{Experimental Setup}

In our study, the original specification documents are provided in PDF format, including a variety of multi-modal content including text, tables, and figures. The signal definition files and the golden RTL designs are formatted in Verilog. To assess the quality of the generated SVAs, we utilize Cadence JasperGold, one of the leading commercial model checking tools. This evaluation leverages the FPV app in JasperGold to ensure a thorough analysis.

Our experimental setup involves the evaluation of three types of LLMs using our developed generation and evaluation methodology:
\begin{enumerate}
    \item GPT-3.5: The freely available commercial version, GPT-3.5 Turbo, supports a context window of up to 16K tokens.
    \item GPT-4o: The state-of-the-art commercial solution, GPT-4o, offers up to 128K token context window and multi-modal capabilities, making it adept at handling the diverse content in the specification documents.
    \item AssertLLM:  Customized GPT-4o by incorporating specialized techniques such as RAG and custom instructions, tailoring the models specialized to the SVA generation task.
\end{enumerate}

\subsection{Evaluation Metrics}

To conduct a thorough evaluation of the generated SVAs, we propose a set of metrics that align with our evaluation methodology. This approach ensures a detailed assessment of the SVAs' quality on both a per-signal and per-design basis.


For each assertion type of an individual signal, our evaluation includes the following metrics: 1) the number of generated SVAs; 2) the number of syntax-correct SVAs; 3) the number of FPV-passed SVAs; 4) COI coverage for all FPV-passed SVAs. We consider an SVA as ``passed'' if the model checker Jaspergold cannot find any counterexample to it
within 5 hours. 
Furthermore, all SVAs are produced directly from LLMs without any subsequent modifications. Once the evaluation for each signal is complete, we aggregate the statistics of the generated SVAs for each design. We then calculate the proportion of these SVAs that are syntactically correct and can pass the FPV checks. Finally, we calculate the COI coverages for all passed SVAs.


\subsection{Assertion Generation Quality}

To illustrate the efficacy of AssertLLM, we apply it to an illustrative design case: the ``I2C'' protocol. The complete specification document for the ``I2C'' design is structured into six main sections, as discussed in Subsection~\ref{sec:nlspec}.
Note that the specification for each signal is unstructured, mainly across the sections like IO ports, registers, and operation.
Additionally, we provide the signal definition file containing the IO ports and architectural registers and all the internal wires and registers defined for detailed RTL implementation.\looseness=-1

The specification for the ``I2C'' design uses    
 natural language to define 23 signals, comprising 17 IO ports and 6 architecture-level registers. For the IO ports, we categorize them into 4 types: clock, reset, control signal, and data signal. The architecture-level registers are similarly categorized, based on their functionality, into control and data types.  Furthermore, the specification for the ``I2C'' design utilizes two waveform diagrams to describe the behaviors on 5 different signals. AssertLLM will extract the described behaviors in the waveforms and generate SVAs for these signals.

\begin{table*}[!h]
\caption{Evaluation of the generated SVAs for design ``I2C''. 
AssertLLM generates 65 properties, with 23 for bit-width, 14 for connectivity, and 28 for function. 86\% of these generated SVAs are evaluated to be correct both syntactically and functionally.}
\resizebox{\textwidth}{!}{

\begin{tabular}{ccc|cccccc}
\hline
\multicolumn{1}{l|}{}                                                & \multicolumn{2}{c|}{}                                        & \multicolumn{4}{c|}{\textbf{\cellcolor[HTML]{C5D9F1}AssertLLM}}                                                                                                                             & \multicolumn{1}{c|}{\cellcolor[RGB]{225, 225, 226}\textbf{GPT-4o}}   & \textbf{GPT-3.5}                                                                                                  \\ \hline
\multicolumn{1}{c|}{}                                                & \multicolumn{2}{c|}{}                                        & \multicolumn{6}{c}{{\color[HTML]{000000} \textbf{Assertion Evaluation (\#. Generated/\#. Syntax Correct/\#. FPV Pass)}}}                                                                                                                                                                                                         \\ \cline{4-9} 
\multicolumn{1}{c|}{}                                                & \multicolumn{2}{c|}{\multirow{-2}{*}{\textbf{Signal Type}}}  & \multicolumn{1}{c|}{\textbf{Width}} & \multicolumn{1}{c|}{\textbf{Conectivity}} & \multicolumn{1}{c|}{\textbf{Function}} & \multicolumn{1}{c|}{\textbf{Signal Total}} & \multicolumn{1}{c|}{\textbf{Function}} &                                                                                                                   \\ \cline{2-8}
\multicolumn{1}{c|}{}                                                & \multicolumn{1}{c|}{}                          & Clock (1)   & \multicolumn{1}{c|}{1/1/1}          & \multicolumn{2}{c|}{/}                                                           & \multicolumn{1}{c|}{1/1/1}                 & \multicolumn{1}{c|}{3/1/0}             &                                                                                                                   \\ \cline{3-8}
\multicolumn{1}{c|}{}                                                & \multicolumn{1}{c|}{}                          & Reset (2)   & \multicolumn{1}{c|}{2/2/2}          & \multicolumn{2}{c|}{/}                                                           & \multicolumn{1}{c|}{2/2/2}                 & \multicolumn{1}{c|}{6/2/0}             &                                                                                                                   \\ \cline{3-8}
\multicolumn{1}{c|}{}                                                & \multicolumn{1}{c|}{}                          & Control (3) & \multicolumn{1}{c|}{3/3/3}          & \multicolumn{1}{c|}{4/4/1}              & \multicolumn{1}{c|}{/}                 & \multicolumn{1}{c|}{7/7/4}                 & \multicolumn{1}{c|}{9/3/0}             &                                                                                                                   \\ \cline{3-8}
\multicolumn{1}{c|}{}                                                & \multicolumn{1}{c|}{\multirow{-4}{*}{IO (17)}} & Data (11)   & \multicolumn{1}{c|}{11/11/11}       & \multicolumn{2}{c|}{/}                                                           & \multicolumn{1}{c|}{11/11/11}              & \multicolumn{1}{c|}{33/11/0}           &                                                                                                                   \\ \cline{2-8}
\multicolumn{1}{c|}{}                                                & \multicolumn{1}{c|}{}                          & Control (2) & \multicolumn{1}{c|}{2/2/2}          & \multicolumn{1}{c|}{10/10/9}            & \multicolumn{1}{c|}{13/13/13}          & \multicolumn{1}{c|}{25/25/24}              & \multicolumn{1}{c|}{6/2/2}             &                                                                                                                   \\ \cline{3-8}
\multicolumn{1}{c|}{\multirow{-8}{*}{\textbf{From natual language}}} & \multicolumn{1}{c|}{\multirow{-2}{*}{Reg (6)}} & Data (4)    & \multicolumn{1}{c|}{4/4/4}          & \multicolumn{1}{c|}{/}                  & \multicolumn{1}{c|}{6/6/4}             & \multicolumn{1}{c|}{10/10/8}               & \multicolumn{1}{c|}{14/4/4}            &                                                                                                                   \\ \cline{1-8}
\multicolumn{3}{c|}{\textbf{From waveform}}                                                                                         & \multicolumn{2}{c|}{/}                                                        & \multicolumn{1}{c|}{9/9/6}             & \multicolumn{1}{c|}{9/9/6}                 & \multicolumn{1}{c|}{4/4/2}             &                                                                                                                   \\ \cline{1-8}
\multicolumn{3}{c|}{}                                                                                                               & \multicolumn{1}{c|}{23/23/23$^\star$}       & \multicolumn{1}{c|}{14/14/10}           & \multicolumn{1}{c|}{28/28/23}          & \multicolumn{1}{c|}{65/65/56}              & \multicolumn{1}{c|}{75/27/8}           &                                                                                                                   \\ \cline{4-8}
\multicolumn{3}{c|}{\multirow{-2}{*}{\textbf{Design Total}}}                                                                        & \multicolumn{1}{c|}{100\%/100\%$^\dagger$}    & \multicolumn{1}{c|}{100\%/71\%}         & \multicolumn{1}{c|}{100\%/82\%}        & \multicolumn{1}{c|}{100\%/86\%}            & \multicolumn{1}{c|}{36\%/11\%}          & \multirow{-10}{*}{\begin{tabular}[c]{@{}c@{}}Can not handle \\ the original \\ specification files.\end{tabular}} \\ \hline
\end{tabular}

}

\begin{tablenotes}\footnotesize
\item $^\star$ The results represent the number of generated SVAs, syntax-correct SVAs, and FPV-passed SVAs.
\item $^\dagger$ The percentages indicate the proportion of syntax-correct SVAs and FPV-passed SVAs.
\end{tablenotes} 

\label{tbl:eval}
\end{table*}

The evaluation of SVAs generated by our AssertLLM is presented in Table~\ref{tbl:eval}. For each signal, we first verify each type of the generated SVAs separately. Then we summarize all the SVAs to provide design-level statistics.
All SVAs related to bit-width checking performed correctly. However, a minor portion of connectivity and function SVAs contained errors. For SVAs generated from natural language, the errors are attributed to misinterpretations of the specification or hallucinations of language model. For SVAs generated from waveform diagrams, the failures are mainly due to AssertLLM's inability to infer behaviors not explicitly depicted in the waveform diagrams, resulting in incomplete assertions that do not pass the FPV check. Overall, 86\% of the SVAs are both syntactically correct and functionally valid.

In addition to assessing AssertLLM's performance, we also conduct an ablation study to compare the SVA generation capabilities of the original GPT-4o and GPT-3.5 models without the additional techniques, demonstrated in Table~\ref{tbl:eval}. When generating SVAs from natural language, the absence of a mechanism for extracting structured signal specifications significantly limits GPT-4o's ability to produce accurate SVAs. Specifically, GPT-4o fails to generate any correct SVAs for I/O ports and only succeeds in creating reset check assertions for registers, resulting in an overall accuracy of just 11\%. Furthermore, when generating SVAs from waveform diagrams, the lack of a specialized waveform analysis method leads to fewer assertions and a lower FPV-passing rate.
As for GPT-3.5, due to the lack of multi-modal processing capabilities, it cannot generate SVAs directly from the original, multi-modal specification files. \looseness = -1

\begin{figure}[!h]
  \centering
  \includegraphics[width=1\linewidth]{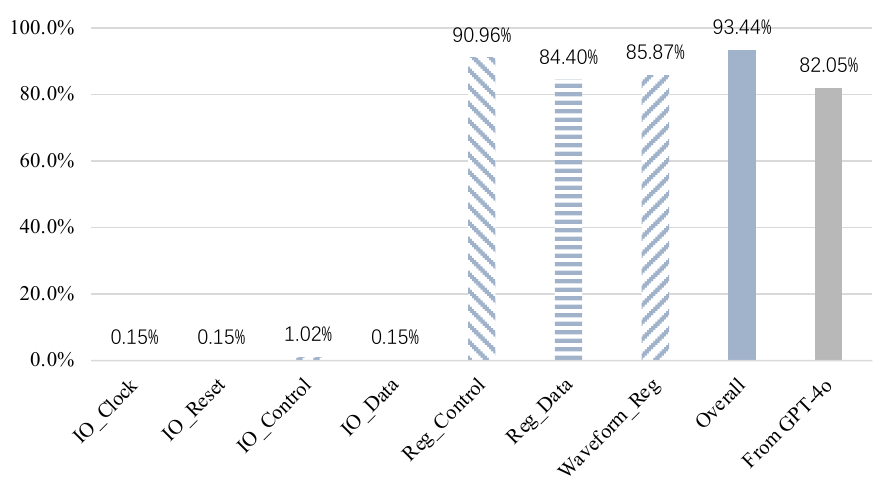}
  \caption{COI coverage of different signal-type SVAs generated by AssertLLM}
  \label{fig:COI}
\end{figure}

We further examine the COI coverage for various signal-type SVAs, as illustrated in Fig.~\ref{fig:COI}. 
Our results demonstrate that SVAs generated by AssertLLM achieve high COI coverage, with a total coverage of 93.44\% (different types of assertions could cover different parts of the design, so the overall rate is higher than individual ones). Additionally, we observe that SVAs generated for registers exhibit higher COI coverage compared to those for IO signals. This can be attributed to the fact that registers typically connect to more logic elements within the design. However, when using GPT-4o to generate SVAs, the COI coverage only reaches 82.05\%, primarily due to its inability to generate a sufficient number of correct SVAs.


\subsection{Assertion Generation for More Designs}
To further analyze the capability of our AssertLLM, we extended its application to generate SVAs for additional designs. In addition to ``I2C,'' we evaluate ``ECG'' and ``Pairing'' designs. The ``ECG'' design calculates the addition of two elements in the elliptic curve group, while the ``Pairing'' design implements Tate bilinear pairing in the elliptic curve group~\cite{frey1994remark}. Table~\ref{tbl:all} presents the overall results, demonstrating AssertLLM's ability to generate high-quality SVAs across various designs. All generated SVAs maintain syntactic correctness and a significant proportion of SVAs can pass FPV. Moreover, the FPV-passed SVAs achieve high COI coverage. These indicate good generalizability of our approach. In contrast, GPT-4o struggles to maintain consistent performance across different designs. It fails to generate any FPV-passed SVAs for the ``ECG'' design and produces only one FPV-passed SVA for the ``Pairing'' design. Consequently, the COI coverage for GPT-4o's SVAs in these two examples is 0\%, as we only calculate COI coverage for FPV-passed assertions.\looseness=-1

\begin{table}[t]
\caption{Evaluation on more designs.}
\resizebox{0.48\textwidth}{!}{

\begin{tabular}{c||c|c||c|c}
\toprule
\multirow{4}{*}{\textbf{Design}} & \multicolumn{4}{c}{\textbf{SVA Evaluation}}                                                                                                                                                                                                                                                                                                   \\ \cline{2-5}
                                 & \multicolumn{2}{c||}{\cellcolor[HTML]{C5D9F1}\textbf{AssertLLM}}                                                                                                                                          & \multicolumn{2}{c}{\cellcolor[RGB]{225, 225, 226}\textbf{GPT-4o}}                                                                                                                                           \\ \cline{2-5}
                                 & \multirow{2}{*}{\begin{tabular}[c]{@{}c@{}}Total   \\      Correctness\end{tabular}} & \multirow{2}{*}{\begin{tabular}[c]{@{}c@{}}Total \\      Coverage\end{tabular}} & \multirow{2}{*}{\begin{tabular}[c]{@{}c@{}}Total \\      Correctness\end{tabular}} & \multirow{2}{*}{\begin{tabular}[c]{@{}c@{}}Total \\      Coverage\end{tabular}} \\ 
                                 &                                                                                      &                                                                                 &                                                                                    &                                                                                 \\ \hline \hline
\multirow{2}{*}{I2C}             & 65/65/56                                                                             & \multirow{2}{*}{93\%}                                                           & 75/27/8                                                                            & \multirow{2}{*}{82\%}                                                           \\ 
                                 & 100\%/86\%                                                                           &                                                                                 & 36\%/11\%                                                                          &                                                                                 \\ \hline
\multirow{2}{*}{ECG}             & 22/22/20                                                                             & \multirow{2}{*}{99\%}                                                           & 11/7/0                                                                              & \multirow{2}{*}{0\%}                                                            \\
                                 & 100\%/91\%                                                                           &                                                                                 & 64\%/0\%                                                                            &                                                                                 \\ \hline
\multirow{2}{*}{Pairing}         & 15/15/14                                                                             & \multirow{2}{*}{100\%}                                                          & 12/8/1                                                                              & \multirow{2}{*}{0\%}                                                            \\
                                 & 100\%/93\%                                                                           &                                                                                 & 67\%/8\%                                                                          &                                                                                 \\ \hline
\textbf{Average}                          & 100\%/90\%                                                                           & 97\%                                                                            & 56\%/6\%                                                                           & 27\%        \\ \bottomrule                                                                   
\end{tabular}

}

\begin{tablenotes}\footnotesize
\item The results share the same format as~\Cref{tbl:eval}.
\end{tablenotes}

\label{tbl:all}
\end{table}

\subsection{Discussion: the Impact of Specification Quality}
The generation of high-quality SVAs from natural language specifications relies not only on the capabilities of LLMs but also on the intrinsic quality of the specification documents themselves. A specification that provides only the basic information of signals, such as their names and simple descriptions, without delving into detailed functionalities or connectivities, inherently limits the potential for generating meaningful SVAs, regardless of the power of the LLMs employed. Conversely, specifications that offer comprehensive details, including clear definitions of signal functionalities and connectivities, can facilitate the generation of SVAs even with relatively simple LLMs. This observation can be concluded in the Table~\ref{tbl:all}. The ``I2C'' specification provides detailed information on registers and functionality, enabling the generation of more SVAs. By contrast, both AssertLLM and GPT-4o can only generate a limited number of SVAs for ``ECG'' and ``Pairing'' due to their less detailed specifications.

\section{Conclusion}
In this paper, we introduce AssertLLM, an automated framework designed for generating assertions from entire specification documents. AssertLLM breaks down this intricate task into three phases: natural language information extraction, waveform description extraction, and assertion generation, leveraging specialized LLMs for each phase. Experimental results show that AssertLLM generates assertions with an average of 88\% passing both syntax-checking and FPV checking. Furthermore, these assertions achieve 97\% COI coverage on average, achieving a notable improvement from GPT-4o and GPT-3.5. \looseness = -1


\begin{acks}
    This work is supported in part by the \grantsponsor{GS501100001809}{National Natural Science Foundation of China}{https://doi.org/10.13039/501100001809} under Grant No.~\grantnum{GS501100001809}{62304194}, by \grantsponsor{GS501100010256}{Guangzhou Municipal Science and Technology Project}{https://doi.org/10.13039/501100010256} under Grant No.~\grantnum{GS501100010256}{2023A03J0013},
    and by Hong Kong Research Grants Council (RGC) ECS under Grant No. 26208723.
\end{acks}

\balance
\bibliographystyle{ACM-Reference-Format}
\bibliography{spec, references_1, references_2, assertion}

\end{document}